\documentclass[aps,pra,nofootinbib,twocolumn]{revtex4-2}

\usepackage{amsmath,amssymb}

\newcommand{\rmi}{{\rm i}}
\newcommand{\rmd}{{\rm d}}
\newcommand{\tr}{{\rm tr}}
\newcommand{\eqtri}{\triangleq}
\newcommand{\ket}[1]{\left|{#1}\right\rangle}
\newcommand{\bra}[1]{\left\langle{#1}\right|}
\newcommand{\braket}[2]{\langle{#1}|{#2}\rangle}
\newcommand{\Dcirc}{\mathcal{D}^{\circ}}

\newcommand{\vlag}[1]{\tau[{#1}]}
\newcommand{\ext}[1]{\varepsilon[{#1}]}
\newcommand{\utag}[1]{\tilde{\tau}[{#1}]}
\newcommand{\uext}[1]{\tilde{\varepsilon}[{#1}]}
\newcommand{\extag}[2]{\varepsilon[{#1}]\tau[{#2}]}
\newcommand{\uextag}[2]{\tilde{\varepsilon}[{#1}]\tau[{#2}]}
\newcommand{\exutag}[2]{\varepsilon[{#1}]\tilde{\tau}[{#2}]}
\newcommand{\uexutag}[2]{\tilde{\varepsilon}[{#1}]\tilde{\tau}[{#2}]}

\begin{document}

\title{Notes on tagged vector spaces}

\author{Filippus S. Roux}
\email{rouxf@ukzn.ac.za}
\affiliation{University of KwaZulu-Natal, Private Bag X54001, Durban 4000, South Africa}

\begin{abstract}
A generalization is provided for the notion of tags, often found in physics notation. The properties of tags and their extractors arediscussed. It leads to the definition of tagged vector spaces. It is applied in the context of quantum optics, to provide a mathematical description for Dirac notation as used in physics. The commutation relations for ladder operators and a symplectic phase space are derived as consequences of the properties of tagged vector spaces. To incorporate the spatiotemporal and spin degrees of freedom, tags are indexed by functions, leading to a functional representation.
\end{abstract}

\maketitle


\section{Introduction}
\label{intro}

In physics, it is convenient to model physical systems in terms of quantities that are invariant with respect to any nonphysical mathematical transformations. When a quantity can change because of such transformations, it becomes necessary to specify the conditions under which a specific representation of the systems by that quantity is relevant. Examples of such transformations are coordinate transformations. Since we are free to use any coordinate system, there are infinitely many ways to represent one and the same object. To avoid this variability, one would use a formalism in which the system is modelled in a way that is invariant to coordinate transformations.

A method to define invariant quantities is to multiply the non-invariant quantity by another auxiliary non-invariant quantity and integrate over the variables that are affected by the transformation. The auxiliary quantity is chosen such that the transformations of the two non-invariant quantities would cancel, leaving the combined quantity invariant.

There are various approaches found in physics to formulate such invariant quantities. However, some cases are more challenging. Often the auxiliary non-invariant quantities are provided by an orthogonal basis. Orthogonality calls for an inner product space \cite{merzbacher,mw}. However, when a suitable orthogonal basis for a particular physical system is continuous, the orthogonality condition implies that the elements of the basis cannot be elements of the inner product space.

Such scenarios are often found in quantum physics when the number of degrees of freedom in the physical system becomes infinite. In this context, a continuous basis is usually presented in physics as the eigenvectors of an unbounded self-adjoint operator $\hat{A}$. The eigenvalue equation is then expressed as
\begin{equation}
\hat{A} \ket{\lambda} = \ket{\lambda} \lambda ,
\end{equation}
where $\lambda$ is the (continuous) eigenvalue, and $\ket{\lambda}$ is its associated eigenvector, expressed in Dirac notation. For the nondegenerate case, the eigenvectors are orthogonal, with an orthogonality condition that reads
\begin{equation}
\braket{\lambda}{\lambda'} = \delta(\lambda-\lambda') ,
\end{equation}
where we ignore any orthogonality constant. The eigenvectors are not normalizable and thus do not belong to the inner product space. While analyses in physics are often based on such continuous bases, they are mathematically awkward.

In mathematics, this situation is avoided by dealing with the eigenvalues only, without involving eigenvectors. The spectral theorem \cite{kreyszig} classifies operators in terms of the nature of the spectrum of eigenvalues for which the resolvent $\hat{R}(\lambda)=\hat{A}-\lambda\mathbb{I}$ is not invertible. Eigenvectors of operators with continuous spectra are relegated to the Schwartz space of tempered distributions, which is part of a rigged Hilbert space (or Gel'fand triplet) \cite{gadella}.

To avoid the mathematical issues associated with eigenvectors of continuous spectra, we need a construction that captures the useful properties in the physics context without incurring the detrimental mathematical issues. Such a construction is based on a generalization of the various forms that are used to define invariant quantities in  physics. In such representations of physical systems, the auxiliary non-invariant quantities are associated with entities that either do not represent any numerical value or for which the numerical value does not carry any meaning in the context in which it is used. Such non-numerical entities serve to combine quantities into one mathematical object, which is invariant to transformations and is more convenient to use in calculations and derivations. The purpose of the non-numerical entities is to label different components of the composite entity and facilitate the extraction of such a component from the composite entity.

\subsection{Current usage}

As illustration, we review some typical examples currently found in physics and mathematics. The first example is the simple case of a three-dimensional vector
\begin{equation}
\mathbf{v} = \vec{e}_x x + \vec{e}_y y + \vec{e}_z z ,
\label{vektor}
\end{equation}
where $\vec{e}_x$, $\vec{e}_y$ and $\vec{e}_z$ are the unit vectors in the $x$-, $y$-, and $z$-directions, respectively, and $x$, $y$, and $z$ represent the numerical values of the three components along these directions. While the unit vectors have specifical physical meanings, they do not represent numerical values. To extract a particular component from the total vector, one performs the dot-product of the total vector with the unit vector along that direction. For example,
\begin{equation}
\vec{e}_x\cdot\mathbf{v} = x. 
\end{equation}

Another example is a \emph{qubit} \cite{nc}, which is a state vector in a two-dimensional Hilbert space. It is often represented in terms of Dirac notation by
\begin{equation}
\ket{\psi} = \ket{0}\alpha + \ket{1}\beta ,
\end{equation}
where $\ket{0}$ and $\ket{1}$ are elements of the two-dimensional basis, and $\alpha$ and $\beta$ are complex coefficients. Normalization requires that $|\alpha|^2+|\beta|^2=1$. The two basis elements are postulated to be normalized $\braket{0}{0}=\braket{1}{1}=1$, and mutually orthogonal $\braket{0}{1}=0$. It thus allows the extraction of a specific component, such as
\begin{equation}
\braket{0}{\psi} = \alpha. 
\end{equation}

In the final example, a \emph{generating function} \cite{as} is formed as a linear combination of an infinite discrete set of functions $\{ g_n(\mathbf{x}) \}$, indexed by integers $n$, and multiplied by an auxiliary variable $\mu$ (serving as the generating parameter) raised to the power of $n$:
\begin{equation}
\mathcal{G}(\mathbf{x},\mu) = \sum_{n=0}^{\infty} \mu^n g_n(\mathbf{x}) . 
\label{gengen}
\end{equation}
The result is one composite function with an additional dependence on the auxiliary variable. Although the auxiliary variable is a numerical variable, its numerical value usually does not carry any meaning in the context in which the generating function is used. The individual functions are extracted from the generating function with the aid of a differential operation
\begin{equation}
\frac{1}{n!} \left. \frac{\partial^n}{\partial \mu^n} 
\mathcal{G}(\mathbf{x},\mu) \right|_{\mu=0} 
= g_n(\mathbf{x}). 
\label{onttrekgen}
\end{equation}

While these examples are different and originate in different subfields of mathematics and physics, they all have some common properties. In all cases, different mathematical entities (components or functions) are combined into a composite entity by labelling the individual entities in some way so that they can be extracted from the composite entity by a suitable operation. This notion is obviously quite powerful, because it reappears in different forms in different contexts.

We refer to the formal devices with which the labelling is done as \emph{tags}. In these examples, the tags are diversely represented either by unit vectors, basis vectors, or auxiliary variables raised to the $n$-th power. In all these cases the tags are indexed by a discrete index. While the first two examples have finite numbers of distinct tags, the last example has a countable infinite number of tags. The tags don't need to represent anything physical in applications of these mathematical scenarios. They can be purely formal devices to alleviate calculation complexity. On the other hand, the index with which these tags are labelled often has a physical meaning, such as the physical directions in space, as found in the first example.

The process whereby the tagged entities are extracted is performed with the aid of \emph{extractors}. In the first example, the extractors are the same as the tags, namely the unit vectors. The second example uses the dual state vectors associated with the basis elements. In the last example, they are provided by the differential process used in (\ref{onttrekgen}). In all these examples, the basic extraction process is defined as an operation that produces the Kronecker delta. For these examples they are
\begin{equation}
\vec{e}_m\cdot\vec{e}_n = \braket{m}{n} 
= \frac{1}{m!} \left. \frac{\partial^m \mu^n}{\partial \mu^m} \right|_{\mu=0} = \delta_{m,n}. 
\end{equation}

\section{Notation and properties}

To generalize these notions, we represent the tags by $\tau_n$ and the extractors by $\varepsilon_n$ where $n$ is an index ranging over the elements in an \emph{index space}. The extraction process is then defined as
\begin{equation}
\varepsilon_m \tau_n = \delta_{m,n} ~~~~ {\rm for} ~~~ m,n \in \mathbb{N} ~ ({\rm or} ~ \mathbb{Z}). 
\label{onttrekd}
\end{equation}
In general, tags can be indexed by any set, depending on the context. It can also be a continuous variable taken from the real numbers $\mathbb{R}$, or even by a function space for the definition of functionals. If the index space is $\mathbb{R}$, the tags and extractors become functions of the continuous index $\tau(\nu)$ and $\varepsilon(\nu)$. The extraction process then leads to a Dirac delta function instead of a Kronecker delta
\begin{equation}
\varepsilon(\mu) \tau(\nu) = \delta(\mu-\nu). 
\label{onttrekc}
\end{equation}

The purpose of tags is to form linear combinations of coefficients. Hence,
\begin{equation}
\mathbf{F} = \sum_{n=0}^{\infty} \tau_n F_n .
\end{equation}
The coefficients $F_n$ are elements in a \emph{coefficient space}. The extraction process produces
\begin{equation}
\varepsilon_m \mathbf{F} = \sum_{n=0}^{\infty} \varepsilon_m \tau_n F_n 
= \sum_{n=0}^{\infty} \delta_{m,n} F_n = F_m. 
\end{equation}

In the continuous case, the linear combinations are integrals ranging over the complete index space with integration boundaries not shown
\begin{equation}
\mathbf{F} = \int \tau(\nu) F(\nu)\ \rmd\nu .
\label{defkompf}
\end{equation}
Here $F(\nu)$ is a \emph{coefficient function}. Due to the tag in the integrand in (\ref{defkompf}), the integral does not evaluate to a numerical value, and the integration measure is purely abstract. However, when an extractor removes the tag and produces the Dirac delta function in the integrand, the integral is associated with a tempered distribution
\begin{align}
\varepsilon(\mu) \mathbf{F} & = \int \varepsilon(\mu)\tau(\nu) F(\nu)\ \rmd\nu \nonumber \\ 
& = \int \delta(\mu-\nu) F(\nu)\ \rmd\nu = F(\mu). 
\end{align}

\subsection{Unitary invariance}

In some cases, the representation of the composite quantity in terms of the tags and their components (or coefficients) is not unique. The same composite object can also be represented as a linear combination of a different set of tags, provided that it is related to the original set of tags by a unitary transformation. Such a different representation is produced by inserting an identity matrix or kernel (a Kronecker delta or a Dirac delta function) decomposed into a unitary matrix or kernel and its Hermitian adjoint. We refer to this property as the \emph{unitary invariance} of the composite entity's representation in terms of tags. The unitary invariance in the (finite) discrete case leads to
\begin{equation}
\mathbf{F} = \sum_n \tau_n F_n = \sum_{m,n,p} \tau_n U_{n,m} U_{m,p}^{\dag} F_p 
= \sum_n \tau_m' F_m'. 
\end{equation}
In the continuous case, it is given by
\begin{align}
\mathbf{F} & = \int \tau(\nu) F(\nu)\ \rmd\nu \nonumber \\
& = \int \tau(\nu) U(\nu,\eta) U^{\dag}(\eta,\mu) F(\mu)\ \rmd\nu\ \rmd\eta\ \rmd\mu \nonumber \\
& = \int \tau'(\eta) F'(\eta)\ \rmd\eta ,
\label{tratoes}
\end{align}
where the \emph{unitary kernels} satisfy the condition
\begin{equation}
\int U(\nu,\eta) U^{\dag}(\eta,\mu)\ \rmd\eta = \delta(\nu-\mu) .
\end{equation}
The transformations of the tags and the coefficient function are respectively given by
\begin{align}
\begin{split}
\tau'(\eta) & = \int \tau(\nu) U(\nu,\eta)\ \rmd\nu , \\
F'(\eta) & = \int U^{\dag}(\eta,\mu) F(\mu)\ \rmd\mu .
\end{split}
\label{travlag}
\end{align}

The notion of unitary invariance is of utmost significance in quantum physics. It provides a method to model quantum systems without the imposition of having to specify the coordinate system. Unitary invariance can be generalized to other forms of coordinates invariance that manifest in various forms in physics formalism. An example is differential geometry \cite{baez} where it facilitates the formulation of a coordinate invariant mathematical structure, which is relevant for various physical scenarios. It also lies at the foundation of the background independence (or diffeomorphism invariance) that is required for the formulation of a quantum theory of gravity \cite{gravth}.

As a general principle in the modelling of physical systems, any mathematical representation that involves specific coordinates (independent variables) is less desirable than a representation that is invariant with respect to such coordinates. The former is only valid under the arbitrary condition of such an assumed coordinate system which has nothing to do with the physical system.

\subsection{Associated and unassociated tags and extractors}

Based on unitary invariance, a scenario in which composite quantities are formed as linear combinations of tags allows infinitely many different possible sets of tags, all related via unitary transformations. It also means that the simple relationships given in (\ref{onttrekd}) and (\ref{onttrekc}) on which the extraction process is based can become more general. For those in (\ref{onttrekd}) and (\ref{onttrekc}), we say that the extractors are \emph{associated} with their tags. For \emph{unassociated} extractors, the more general operations produce
\begin{equation}
\varepsilon_m \tau_n' = U_{m,n} ~~~ {\rm or} ~~~ \varepsilon_m' \tau_n = U_{m,n}^{\dag} ,
\label{onttrekud}
\end{equation}
in the finite discrete case, and
\begin{equation}
\varepsilon(\mu) \tau'(\nu) = U(\mu,\nu) ~~~ {\rm or} ~~~
\varepsilon'(\mu) \tau(\nu) = U^{\dag}(\mu,\nu) ,
\label{onttrekuc}
\end{equation}
in the continuous case. To see how this general extraction process works, we start with the last expression in (\ref{tratoes}), using (\ref{travlag}), and apply $\varepsilon(\mu)$ on it, leading to
\begin{align}
\varepsilon(\mu) \mathbf{F} & = \int \varepsilon(\mu) \tau'(\nu) F'(\nu)\ \rmd\nu \nonumber \\
& = \int \varepsilon(\mu) \tau(\eta) U(\eta,\nu) F'(\nu)\ \rmd\eta\ \rmd\nu \nonumber \\
& = \int \delta(\mu-\eta) U(\eta,\nu) F'(\nu)\ \rmd\eta\ \rmd\nu \nonumber \\
& = \int U(\mu,\nu) F'(\nu)\ \rmd\nu . 
\end{align}
A comparison of the second and last expressions demonstrates (\ref{onttrekuc}). The extractors are also related by unitary transformations. Compared to (\ref{travlag}), they transform as
\begin{equation}
\varepsilon'(\eta) = \int U^{\dag}(\eta,\nu) \varepsilon(\nu)\ \rmd\nu. 
\label{traekst}
\end{equation}

The unitary kernels in (\ref{onttrekuc}) are in general tempered distributions. Bare extractors seldomly operate on bare tags. Instead, extractors operate on linear combinations of tags as in (\ref{defkompf}), leading to suitable definitions for the distributions thus produced.

\subsection{Unbiased tags and extractors}

There are special cases of unassociated tags and extractors that are referred to as being \emph{unbiased}. A set of tags and a set of extractors are called (mutually) unbiased when the magnitude of the result of a bare extraction process between them is independent of their respective indices. Hence, $|\tilde{\varepsilon}(\mu)\tau(\nu)|=c$, 
where $\{\tilde{\varepsilon}\}$ is a set of extractors that is unbiased with respect to a set of tags $\{\tau\}$, and $c$ is a constant independent of $\mu$ and $\nu$. It implies that
$\tilde{\varepsilon}(\mu)\tau(\nu)=c\exp\left[\rmi\phi(\mu,\nu)\right]$,
where $\phi(\mu,\nu)$ is a real monotonic (phase) function of $\mu$ and $\nu$.

A ubiquitous example is where this complex phase factor is the \emph{Fourier kernel}
\begin{equation}
\tilde{\varepsilon}(\mu) \tau(\nu) = \exp(-\rmi\mu\nu) . 
\end{equation}
In what follows, we always assume that unbaised sets of tags and extractors produce the Fourier kernel. When $\tilde{\varepsilon}$ is applied to a linear combination of tags defined in terms of a coefficient function $F(\nu)$, the result is the Fourier transform of $F(\nu)$:
\begin{align}
\tilde{\varepsilon}(\mu) \mathbf{F} & = \int \tilde{\varepsilon}(\mu) \tau(\nu) F(\nu)\ \rmd\nu \nonumber \\
& = \int \exp(-\rmi\mu\nu) F(\nu)\ \rmd\nu \equiv \mathcal{F}\{F\}(\mu). 
\end{align}

\subsection{Adjoint composite entity}

The extractor is always placed on the left-hand side of the tag on which it operators, so that the extractor always operates toward the right. However, we can also regard it as a process where the tag ``operates'' toward the left on the extractor. So, with the extactor placed on the left and the tag on the right for an extraction process, the identities of the ``operator'' and the ``operand'' is ambiguous. The roles of the tag and the extractor can be exchanged. The latter can be combined into an \emph{adjoint} composite entity, represented as
\begin{equation}
\mathbf{G}^{\dag} = \int G^*(\mu) \varepsilon(\mu)\ \rmd\mu ,
\end{equation}
where $G^*(\mu)$ is the coefficient function. Applying it to a tag on its right, we get
\begin{align}
\mathbf{G}^{\dag} \tau(\nu) & = \int G^*(\mu) \varepsilon(\mu) \tau(\nu)\ \rmd\mu \nonumber \\
& = \int G^*(\mu) \delta(\mu-\nu)\ \rmd\mu = G^*(\nu). 
\end{align}
In the process, the tag ``extracted'' the coefficient of the extractor from the composite entity. Note that the order of the extractor and the tag remains the same: extractor on the left and tag on the right.

\subsection{Zipping and unzipping}

When both the tags and the extractors are combined into respective composite entities, they can operate on each other to produce a single summation of integrations over the coefficients of the two respective objects
\begin{align}
\mathbf{G}^{\dag} \mathbf{F} & = \int G^*(\mu) \varepsilon(\mu) \tau(\nu) F(\nu)\ \rmd\nu\ \rmd\mu \nonumber \\
& = \int G^*(\mu) \delta(\mu-\nu) F(\nu)\ \rmd\nu\ \rmd\mu \nonumber \\
& = \int G^*(\nu) F(\nu)\ \rmd\nu . 
\label{zip}
\end{align}
This process is called \emph{zipping} because it multiplies the coefficients of the two coefficient functions for the same index value under one integral.

One can also have the opposite situation where an integration of the product of two functions is \emph{unzipped} (separated) into separate composite entities with these functions as the respective coefficient functions. It is done by inserting the identity kernel resolved in terms of tags and extractors between the two functions. The procedure is represented by (\ref{zip}) in reversed order.

\subsection{Identity and dressed kernels}

Based on the order in which extractors and tags are written for an extraction process, we can write them in the opposite order as $\tau(\nu)\varepsilon(\mu)$ when we do not want them to perform an extraction process. The inverse order allows us to resolve an \emph{identity operator} $\mathbb{I}$ in terms of tags and their associated extractors, integrated (or summed) over the entire index space
\begin{equation}
\mathbb{I} \eqtri \int \tau(\nu) \varepsilon(\nu)\ \rmd\nu \equiv \int \tau(\nu)
\delta(\nu-\mu) \varepsilon(\mu)\ \rmd\nu\ \rmd\mu. 
\end{equation}
It follows that $\mathbb{I}\tau=\tau$ and $\varepsilon\mathbb{I}=\varepsilon$. A generalization of the identity operator leads to a \emph{dressed kernel}
\begin{equation}
\hat{A} \eqtri \int \tau(\nu) A(\nu,\mu) \varepsilon(\mu)\ \rmd\nu\ \rmd\mu ,
\label{kwantop}
\end{equation}
where the hat $\hat{\cdot}$ indicates that a kernel $A(\mu,\nu)$ is \emph{dressed} on the left by a tag and on the right by an extractor. The term ``dressed'' implies an integration (or summation) over a shared index. Below, it is used to define operators.

\section{\label{deftvs}Tagged vector spaces}

The above discussion of the notation and properties of tags and extractors serves as background for the 
definition of \emph{tagged vector spaces}. For this purpose, we need to specify the \emph{index space} $\mathcal{I}$ and the \emph{coefficient space} $\mathcal{C}$. The elements of the index space are used to label the tags and the extractors. It remains the same even when we transform the tags or extractors via unitary transformations. Here, we assume a continuous index space (such as $\mathbb{R}$). The elements of the coefficient space provide the coefficients in the linear combinations of tags or extractors. Since the labels of the coefficients must be the same as those of the tags, the elements of the coefficient space are functions of the index space
\begin{equation}
\mathcal{C} \eqtri \{ F : \mathcal{I} \rightarrow \mathcal{T} \}. 
\end{equation}
The range $\mathcal{T}$ (or target space) is usually either $\mathbb{R}$ or $\mathbb{C}$, but can also be more complex.

The set of tags $\{\tau\}$ and the set of their associated extractors $\{\varepsilon\}$, both labelled by the same index space, are defined with (at least) these two postulates:
\begin{itemize}
\item \textbf{orthogonality}:
\begin{equation}
\varepsilon(\mu)\tau(\nu) = \delta(\mu-\nu) ,
\end{equation}
\item \textbf{completeness}:
\begin{equation}
\int \tau(\nu) \varepsilon(\nu)\ \rmd\nu=\mathbb{I}. 
\end{equation}
\end{itemize}

For a given set of tags $\{\tau\}$, the tagged vector space is defined as
\begin{equation}
\Upsilon(\mathcal{I},\mathcal{C}) \eqtri \left\{ \mathbf{F}=\int \tau(\nu) F(\nu)\ \rmd\nu :
\nu\in\mathcal{I}, F\in\mathcal{C} \right\} ,
\label{vlgruimte}
\end{equation}
Unitary invariance allows the elements of this tagged vector space to be expressed in terms of any other set of tags 
$\{\tau'\}$ that is unitarily related to $\{\tau\}$.

Every tagged vector space is associated with an \emph{adjoint tagged vector space}, defined in terms of the same \emph{index space} $\mathcal{I}$ and a \emph{coefficient space} $\mathcal{C}'$ that may be the same as that of the tagged vector space. (Scenarios where $\mathcal{C}'\neq\mathcal{C}$ may be rare, but can be useful in certain situations.) The linear combinations in the adjoint tagged vector space are formed with the associated set of extractors $\{\varepsilon\}$. Hence,
\begin{equation}
\Upsilon' \eqtri \left\{ \mathbf{G}^{\dag}=\int G^*(\mu)\varepsilon(\mu)\ \rmd\mu :
\mu\in\mathcal{I}, G^*\in\mathcal{C}' \right\} ,
\label{advlgruimte}
\end{equation}
Again, unitary invariance allows elements of the adjoint tagged vector space to be expressed in terms of any other set of extractors $\{\varepsilon'\}$ unitarily related to $\{\varepsilon\}$. 
In the rest of this paper, we'll assumed that $\mathcal{C}'\equiv\mathcal{C}$.

\subsection{Operators}

Operators on tagged vector spaces are defined as dressed kernels (\ref{kwantop}). When an operator is applied to a composite entity (a linear combination of tags), it produces a zipping process, leading to
\begin{align}
\hat{A} \mathbf{F} & = \int \tau(\nu) A(\nu,\mu) \varepsilon(\mu) 
\tau(\xi) F(\xi)\ \rmd\nu\ \rmd\mu\ \rmd\xi \nonumber \\
& = \int \tau(\nu) A(\nu,\mu) F(\mu)\ \rmd\nu\ \rmd\mu \nonumber \\
& = \int \tau(\nu) F'(\nu)\ \rmd\nu = \mathbf{F}' ,
\label{afnaf}
\end{align}
where
\begin{equation}
F'(\nu) = \int A(\nu,\mu) F(\mu)\ \rmd\mu ,
\end{equation}
is a transformed coefficient function, as produced by the action of the operator. The same operator can also be applied to an adjoint composite entity (a linear combination of extractors). It then operates toward the left via a similar zipping process to give
\begin{align}
\mathbf{G}^{\dag} \hat{A} & = \int G^*(\xi) \varepsilon(\xi) \tau(\nu) A(\nu,\mu) 
\varepsilon(\mu)\ \rmd\nu\ \rmd\mu\ \rmd\xi \nonumber \\
& = \int G^*(\nu) A(\nu,\mu) \varepsilon(\mu)\ \rmd\nu\ \rmd\mu \nonumber \\
& = \int G^{\prime *}(\mu) \varepsilon(\mu)\ \rmd\mu = \mathbf{G}^{\prime\dag} .
\label{ganag}
\end{align}
Here
\begin{equation}
G^{\prime *}(\mu) = \int G^*(\nu) A(\nu,\mu)\ \rmd\nu ,
\end{equation}
is the transformed coefficient function of the adjoint composite entity. It follows that operators, defined as dressed kernels, can operate in either direction with impunity.

When an operator is sandwiched between both types of linear combinations, the order of evaluation is unimportant. Two zipping processes are performed to produce
\begin{equation}
\mathbf{G}^{\dag} \hat{A} \mathbf{F} = \int G^*(\nu) A(\nu,\mu) F(\mu)\ \rmd\nu\ \rmd\mu ,
\end{equation}
for any order of integration; the process is \emph{associative}.

\subsection{Adjoint}

The (Hermitian) \emph{adjoint} is a bijective mapping between $\Upsilon$ and $\Upsilon'$, defined by
\begin{align}
\dag : \Upsilon(\mathcal{I},\mathcal{C}) & \longrightarrow \Upsilon'(\mathcal{I},\mathcal{C}) \nonumber \\
\mathbf{F} = \int \tau(\nu) F(\nu)\ \rmd\nu & \longmapsto \mathbf{F}^{\dag} = \int F^*(\nu) \varepsilon(\nu)\ \rmd\nu ,
\end{align}
where $F^*$ is the complex conjugate of $F$. The adjoint replaces tags by their associated extractors and the coefficient functions by their complex conjugates.

The same process can be performed in reverse
\begin{align}
\dag : \Upsilon'(\mathcal{I},\mathcal{C}) & \longrightarrow 
\Upsilon(\mathcal{I},\mathcal{C}) \nonumber \\
\mathbf{F}^{\dag} = \int F^*(\mu) \varepsilon(\mu)\ \rmd\mu
& \longmapsto \mathbf{F} = \int \tau(\mu) F(\mu)\ \rmd\mu ,
\end{align}
which implies that $\mathbf{F}^{\dag\dag} \equiv \mathbf{F}$. It provides an equivalent mapping (an involution) from $\Upsilon'$ to $\Upsilon$ in which extractors are replaced by their associated tags, and coefficient functions by their complex conjugates.

When applied to operators, the mapping produces $\hat{A} \longmapsto \hat{A}^{\dag}$, where
\begin{align}
\begin{split}
\hat{A} & = \int \tau(\nu) A(\nu,\mu)\ \varepsilon(\mu)\ \rmd\nu\ \rmd\mu , \\
\hat{A}^{\dag} & = \int \tau(\mu) A^{\dag}(\mu,\nu)\ \varepsilon(\nu)\ \rmd\nu\ \rmd\mu .
\end{split}
\end{align}
The adjoint kernel $A^{\dag}(\mu,\nu)$ is the transposed complex conjugate of the kernel $A(\nu,\mu)$. The tags and extractors are replaced by each other and their order is inverted. When the adjoint mapping is applied to a product of operators, the order of the operators is likewise inverted. The same applies when composite entities are involved. For example.
\begin{equation}
\left(\hat{A} \mathbf{F}\right)^{\dag} = \mathbf{F}^{\dag} \hat{A}^{\dag} ~~~ {\rm and} ~~~
\left(\mathbf{F}^{\dag} \hat{A}\right)^{\dag} = \hat{A}^{\dag} \mathbf{F} . 
\end{equation}
When $A^{\dag}(\mu,\nu)=A(\nu,\mu)$, the operator is \emph{self-adjoint}, provided that the domains of the operators are the same.

\subsection{Operator trace}

In terms of tags and extractors, an \emph{operator trace} is performed by moving the (left-most) tag to the right-hand side of the (right-most) extractor, which then produces the Dirac delta function. (In case the tag and extractor are not associated, the appropriate unitary kernel is produced instead of the Dirac delta function.) Hence,
\begin{align}
\tr\{\hat{A}\} & \eqtri \tr\left\{\int \tau(\nu) A(\nu,\mu) \varepsilon(\mu)\ \rmd\nu\ \rmd\mu \right\} \nonumber \\
& = \int A(\nu,\mu) \varepsilon(\mu) \tau(\nu)\ \rmd\nu\ \rmd\mu \nonumber \\
& = \int A(\nu,\mu) \delta(\mu-\nu)\ \rmd\nu\ \rmd\mu \nonumber \\
& = \int A(\nu,\nu)\ \rmd\nu. 
\label{operspoor}
\end{align}
The operator trace is converted into a \emph{kernel trace}. In the case of multiple operators, the extractor of a preceeding operator and the tag of a subsequent operator produce a Dirac delta function so that all tags and extractors are removed in the trace process.

The operator trace thus defined involves the entire space of all possible linear combinations of the tags. As a result, it tacitly assumes that the operators are defined on that entire space. Often such traces of operators on that space are divergent. Traces that produce divergent results are not considered to be physically meaningful. In the case where tags are used to model physical scenarios, the domain on which operators are defined is restricted by the physical entities on which they operate. In such cases, traces would normally include such entities so that they evaluate to finite results.

\section{Application: Dirac notation for quantum optics}
\label{appdirac}

The formalisms that were developed during the advent of quantum physics took different paths for those developed by physicists \cite{diracboek,merzbacher,itzykson,peskin,weinberg1,weinberg2}, and those developed in mathematics 
\cite{kreyszig,haagboek,hassani,szekeres}. While the mathematical formulation is suitably rigorous, it represents the usage in a way that differs from what was intended when the original physics formalisms were invented 
\cite{verras,klaczynski}. For example, the idea of a Dirac state vector \cite{diracboek,merzbacher}, as represented by a ket $\ket{\psi}$, is that it is completely agnostic about any coordinate system. Yet, in terms of the current approach in mathematical physics, a ket is an element of a Hilbert space for which the space of square integrable functions $L^2$ is often taken as a suitable representation. A square integrable function begs the question of explicit dependencies on some independent variables. In the physics context, such independent variables carry physical meaning, for example, as time and spatial coordinates or their associated Fourier domain variables. Different coordinate systems can be defined with the aid of coordinate transformation (unitary transformations in the quantum context). With such freedom in defining the coordinate system, all related by unitary transformations, different functions in $L^2$ can represent the same physical state vector, provided that the relationships among their independent variables are taken into account. The opposite situation is where one and the same function in $L^2$ can represent different physical state vectors when the independent variables represent different physical coordinate systems. Without a clear statement of the physical nature of the independent variables a square integrable functions in $L^2$ does not provide a useful model for a physical quantum state. Instead, a coordinate invariant (unitary invariant) representation provides a more valid (and useful) physical representation of such a state.

This invariance is captured in the original formulation of Dirac state vectors in terms of kets. It is accomplished by multiplying the function by a nominal coordinate basis corresponding to the coordinates represented by the independent variables of the function and integrating over the independent variables. For a single-particle state, it is symbolically expressed as
\begin{equation}
\ket{\psi} \eqtri \int \ket{\mathbf{x}} \psi(\mathbf{x})\ \rmd\mathbf{x} ,
\end{equation}
where we leave the careful definition of the independent variables, coordinate basis elements (here shown as kets), and the integration measure unspecified for the time being. Dual state vectors are represented by bras $\bra{\psi}$ in a one-to-one association with their corresponding kets, leading to a symmetric formulation. The inner product is then expressed by $\braket{\psi}{\psi'}$, producing a complex numerical value as result.

The Dirac formalism also allows more freedom in the usage of operators. While the application of an operator to a ket is represented as $\hat{A}\ket{\psi}=\ket{\phi}$, operating toward the right, the Hermitian adjoint of this equation produces the equivalent operation expressed in terms of the bras $\bra{\psi}\hat{A}^{\dag}=\bra{\phi}$. As a result, an operator can also be applied to the bras, operating toward the left.

While the Dirac formalism works well within the quantum physics context, it does not conform to mathematical 
practise \cite{verras}. (See also \cite{klaczynski}, and references therein.) Below, we show how a tagged vector space can reproduce the formalism as used by the physics community while being mathematically rigorous.

\subsection{Bras and kets}

Here, the index space is the real numbers $\mathcal{I}\equiv\mathbb{R}$ and the coefficient space is the space of one-dimensional normalized complex Schwartz functions 
\begin{equation}
\mathcal{C}\eqtri \left\{ \psi\in \mathcal{S}(\mathbb{R},\mathbb{C}) : \|\psi\|=1 \right\} .
\label{koefdef}
\end{equation}
The notation for the tags and extractors is changed to conform to the Dirac notation. Instead of $\tau(\nu)$, we represent the tags as kets $\ket{\nu}\equiv\tau(\nu)$ and their associated extractors as bras $\bra{\mu}\equiv\varepsilon(\mu)$. These tags and extractors do not represent normalized (or normalizable) state vectors on their own. The postulates are
\begin{itemize}
\item \textbf{orthogonality}:
\begin{equation}
\braket{\mu}{\nu} = \delta(\mu-\nu) ,
\end{equation}
\item \textbf{completeness}:
\begin{equation}
\int \ket{\nu} \bra{\nu}\ \rmd\nu = \mathbb{I}. 
\end{equation}
\end{itemize}
The tagged vector space and the adjoint tagged vector space are defined as
\begin{align}
\begin{split}
\Upsilon & \eqtri \left\{\ket{\psi} = \int \ket{\nu} \psi(\nu)\ \rmd\nu : \nu\in\mathbb{R}, 
\psi\in \mathcal{C} \right\} , \\
\Upsilon' & \eqtri \left\{\bra{\phi} = \int \phi^*(\mu) \bra{\mu}\ \rmd\mu : \mu\in\mathbb{R}, 
\phi^*\in \mathcal{C} \right\} . 
\end{split}
\label{vlgkwa}
\end{align}
We also define an \emph{inner product}, represented as 
\begin{align}
\braket{\phi}{\psi} & = \int \phi^*(\mu) \braket{\mu}{\nu}\psi(\nu)\ \rmd\mu\ \rmd\nu \nonumber \\
& = \int \phi^*(\mu) \delta(\mu-\nu)\psi(\nu)\ \rmd\mu\ \rmd\nu \nonumber \\
& = \int \phi^*(\nu)\psi(\nu)\ \rmd\nu ,
\end{align}
equivalent to an inner product between the coefficient functions. The tagged vector space thus leads to an \emph{inner product space},\footnote{We do not convert the inner product space to a complete Hilbert space with a coefficient space given by $L^2(\mathbb{R})$ because the coefficient space needs to be $\mathcal{S}(\mathbb{R})$ to represent physical states.} with elements given by normalized \emph{state vectors} denoted by kets $\ket{\psi}$.

Although the inner product between state vectors corresponds to an inner product between their coefficient functions, implying an isomorphism between the tagged vector space and the coefficient space, they do not \emph{physically} represent the same space. The unitary invariant definition of state vectors, disallow a unique mapping between state vectors in the inner product space and the elements of the coefficient space. All representations of $\ket{\psi}$ (or $\bra{\phi}$) in terms of different tags (or extractors) via unitary transformations represent the same physical state.

The normalization condition in (\ref{koefdef}) is expressed by the \emph{Euclidean norm}. As a result, $\ket{\psi}$ is a \emph{normalized} state vector in that $\|\psi\|^2=\braket{\psi}{\psi}=1$. By implication, the coefficient functions are normalized. Similarly, $\bra{\phi}$ represents a normalized adjoint state vector. The norm in turn implies a metric given by the \emph{Euclidean metric} which leads to an associated metric topology.

\subsection{Operators}

For the sake of the following discussion, we only consider operators that are defined on the entire tagged vector space and the entire adjoint tagged vector space.\footnote{Since these spaces are not Hilbert spaces, these operators include unbounded operators.} We also assume that these operators map the elements of these spaces back to themselves, albeit with possible numerical factors. Such numerical factors are allowed provided that we relax the restriction on normalization by allowing a subsequent normalization process, so that the operators may include those that are not \emph{trace preserving}. For an operator $\hat{A}$, we have
\begin{equation}
\hat{A} : \Upsilon \longrightarrow \Upsilon ~~~ {\rm and} ~~~
\hat{A} : \Upsilon' \longrightarrow \Upsilon'. 
\label{opmap}
\end{equation}
Stated differently, $\ket{\psi}\in\Upsilon$ if and only if $\hat{A}\ket{\psi}\in\Upsilon$, and $\bra{\phi}\in\Upsilon'$ if and only if $\bra{\phi}\hat{A}\in\Upsilon'$. To justify this restriction, we note that an operator representing a physical process must be well-defined for all physical states. Since all the elements of the (adjoint) tagged vector space represent physical states, such operators must be defined on the entire (adjoint) tagged vector space.

These operators include \emph{density operators}. For a pure state, such a density operator is defined as
\begin{equation}
\hat{\rho}_{{\rm pure}} \eqtri \ket{\psi}\bra{\psi} \equiv \int \ket{\nu} 
\psi(\nu)\psi^*(\mu) \bra{\mu}\ \rmd\nu\ \rmd\mu ,
\end{equation}
in terms of (\ref{kwantop}). To generalize it to include mixed states, we also allow ensemble averages of such pure states, leading to the general definition
\begin{equation}
\hat{\rho} \eqtri \int \ket{\nu} \rho(\nu,\mu) \bra{\mu}\ \rmd\nu\ \rmd\mu ,
\end{equation}
where $\rho(\nu,\mu)$ represents a self-adjoint positive semi-definite Hilbert-Schmidt kernel. The \emph{purity} of the state is given by its Hilbert-Schmidt norm. The normalization of the density operator requires that its operator trace equals 1. According to (\ref{operspoor}), it implies that
\begin{align}
\tr\{\hat{\rho}\} & = \int \rho(\nu,\mu) \braket{\mu}{\nu}\ \rmd\nu\ \rmd\mu \nonumber \\
& = \int \rho(\nu,\nu)\ \rmd\nu = 1. 
\end{align}

General operators are defined as in (\ref{kwantop}), given by
\begin{equation}
\hat{A} \eqtri \int \ket{\nu} A(\nu,\mu) \bra{\mu}\ \rmd\nu\ \rmd\mu ,
\end{equation}
in terms of our current notation. The kernel $A(\nu,\mu)$ of a general operator need not be a Hilbert-Schmidt kernel. In general, these kernels form a subset of the \emph{Schwartz kernels}. By relaxing the restriction on normalization, we include those kernels that are not trace preserving.

The operations of these operators on the kets and bras in terms of the tags and extractors are represent by (\ref{afnaf}) and (\ref{ganag}) when using kets and bras for the tags and extractors, respectively. When the adjoint is applied to such a process, we get
\begin{equation}
\bra{\psi'} = \left(\ket{\psi'}\right)^{\dag} = \left(\hat{A} \ket{\psi}\right)^{\dag} 
= \bra{\psi} \hat{A}^{\dag}. 
\end{equation}
As with $\hat{A}$, the adjoint $\hat{A}^{\dag}$ can operate toward either the right or the left. Moreover,
\begin{equation}
\ket{\phi'} = \left(\bra{\phi'}\right)^{\dag} = \left(\bra{\phi}\hat{A}\right)^{\dag} 
= \hat{A}^{\dag} \ket{\phi}. 
\end{equation}

In the tagged vector space formulation, operators are \emph{never} placed inside kets or bras because that would lead to confusion. The symbol inside kets and bras are used as labels to distinguish them and do not represent anything on which an operator can operate (the operator does not directly operate on the coefficient function). Although we generally use the same symbol for the coefficient function to label the state vector, this is not always the case. For example, the kets and bras representing Fock states are usually labelled by the occupation number. Operators are defined only to operate on state vectors and adjoint state vectors. As such, the operator is always placed outside the ket or the bra on which it operates on the side of the straight edge: $\hat{A}\ket{\psi}$ or $\bra{\psi}\hat{A}$. To avoid possible confusion, quantities that are not operators can be placed on the side of the pointed edge: 
$\ket{\psi}f$ or $f\bra{\psi}$, but it is not a strict rule.

\subsection{Probability distributions and their moments}
\label{eenmomop}

The normalization imposed on the state vectors is required for the conservation of probability as incorporated into the formalisms of quantum theory. In the tagged vector space formulation, it implies that the modulus squared coefficient function can serve as a probability distribution
$|\psi(\nu)|^2\equiv P(\nu)$, so that
\begin{equation}
\int |\psi(\nu)|^2\ \rmd\nu \equiv \int P(\nu)\ \rmd\nu = 1. 
\end{equation}

The moments of such a probability distribution are often found in measurements or observations. These moments are defined by
\begin{equation}
\mathcal{M}_n \eqtri \int \nu^n P(\nu)\ \rmd\nu \equiv \int \nu^n |\psi(\nu)|^2\ \rmd\nu ,
\end{equation}
where $n$ represents the order of the moment. It is convenient to compute such moments using tags and extractors in \emph{moment operators} defined as diagonal operators
\begin{equation}
\hat{M}_n \eqtri \int \ket{\nu} \nu^n \bra{\nu}\ \rmd\nu . 
\end{equation}
A moment operator, sandwiched between the bra and ket of the same state, produces
\begin{align}
\bra{\psi}\hat{M}_n\ket{\psi}
& = \int \psi^*(\mu) \braket{\mu}{\nu} \nu^n \braket{\nu}{\xi} \psi(\xi)\ \rmd\mu\ \rmd\nu\ \rmd\xi \nonumber \\
& = \int \nu^n |\psi(\nu)|^2\ \rmd\nu \equiv \mathcal{M}_n. 
\end{align}
The product of any two moment operators gives another moment operator with the sum of their orders 
$\hat{M}_m\hat{M}_n=\hat{M}_{m+n}$. It thus suffices to define a \emph{first-order moment operator}. All the higher order moments can then be produced as
\begin{equation}
\bra{\psi}\hat{M}_1^n\ket{\psi} = \int \psi^*(\nu) \nu^n \psi(\nu)\ \rmd\nu \equiv \mathcal{M}_n. 
\end{equation}

Provided that these moment operators are only applied to state vectors whose coefficient functions are Schwartz functions, these moments are always finite and therefore defined on the entire tagged vector space. Although finite, they can take on any value and are therefore \emph{unbounded}.

A first-order moment operator assumes a specific representation of the tags as labelled by the index space. If we apply a unitary transformation to define the operator in terms of a different set of tags, the result is not a first-order moment operator for the new set of tags. So, for a different unitarily related set of tags and extractors
\begin{align}
\begin{split}
\ket{\mu}' & = \int \ket{\nu} U(\nu,\mu)\ \rmd\nu , \\
\bra{\mu}' & = \int U^{\dag}(\mu,\xi) \bra{\xi}\ \rmd\xi ,
\end{split}
\end{align}
the transformed version of the first-order moment operator is not a first-order moment operator for $\mu$, because
\begin{equation}
\int U^{\dag}(\mu,\nu) \nu U(\nu,\xi)\ \rmd\nu \neq \mu. 
\end{equation}
To compute the first-order moments with respect to a different set of tags that is unitarily related to the original set of tags, a dedicated first-order moment operator must be defined for that different set of tags. Often, separate first-order moment operators are defined for two mutually unbaised sets of tags.

\subsection{Quadrature operators}

An example of such a pair of first-order moment operators, is the \emph{quadrature operators}, often encountered in quantum optics. To define quadrature operators, we use two different mutually unbiased sets of tags (and extractors). The indices for these two sets of tags are labelled by the variables $q\in\mathbb{R}$ and $p\in\mathbb{R}$, respectively, so that the tags are indicated as $\ket{q}$ and $\ket{p}$, respectively, and their associated extractors by $\bra{q}$ and $\bra{p}$, respectively. The postulates then read
\begin{itemize}
\item \textbf{orthogonality}:
\begin{equation}
\braket{q}{q'} = \delta(q-q') ~~~ {\rm and} ~~~ \braket{p}{p'} = 2\pi\delta(p-p') ,
\label{ax1}
\end{equation}
\item \textbf{completeness}:
\begin{equation}
\int \ket{q}\bra{q}\ \rmd q = \mathbb{I} ~~~ {\rm and} ~~~ 
\int \ket{p}\bra{p}\ \frac{\rmd p}{2\pi} = \mathbb{I} ,
\label{ax2}
\end{equation}
\item \textbf{unbiased}:
\begin{equation}
\braket{q}{p} = \exp(\rmi qp) ~~~ {\rm and} ~~~ \braket{p}{q} = \exp(-\rmi qp) .
\label{ax3}
\end{equation}
\end{itemize}
The third set of postulates are included for the mutually unbiased property. The postulates for the $p$ index space incorporate factors of $2\pi$ to anticipate the definition of the Fourier transform. The \emph{quadrature operators} are 
defined as first-order moment operators
\begin{equation}
\hat{q} \eqtri \int \ket{q} q \bra{q} \rmd q ~~~ {\rm and} ~~~ \hat{p} \eqtri \int \ket{p} p \bra{p} \frac{\rmd p}{2\pi} .
\label{kwadef}
\end{equation}
The domains for these operators are as expressed in (\ref{opmap}) in terms of the spaces in (\ref{vlgkwa}).

\subsection{Quadrature operator commutation}
\label{kwadkomm}

Considering the commutation relations of the quadrature operators, we can avoid having to define one of them in terms of a derivative, which would have implied that it could only operate toward the right, thanks to their tagged vector space definitions. The tags and extractors allow them to operate toward either direction.

For the commutation $[\hat{q},\hat{p}]$ in terms of (\ref{kwadef}), we obtain
\begin{align}
[\hat{q},\hat{p}] = & \int \ket{q} q \braket{q}{p} p \bra{p} \rmd q \frac{\rmd p}{2\pi}
- \int \ket{p} p \braket{p}{q} q \bra{q} \frac{\rmd p}{2\pi} \rmd q \nonumber \\
= & \int \ket{q} q \exp(\rmi qp) p \bra{p} \rmd q \frac{\rmd p}{2\pi} \nonumber \\
& - \int \ket{p} p \exp(-\rmi qp) q \bra{q} \frac{\rmd p}{2\pi} \rmd q ,
\end{align}
where we used (\ref{ax3}). To make the two terms more compatible, we insert identities resolved in a $q$-basis, as in (\ref{ax2}), on the side of $\hat{p}$ in each term, and relabel the $q$-indices to bring the two terms into the same form. The result reads
\begin{align}
[\hat{q},\hat{p}] = & \int \ket{q} q \exp(\rmi qp) p \exp(-\rmi q'p) 
\bra{q'} \rmd q \frac{\rmd p}{2\pi} \rmd q' \nonumber \\
& - \int \ket{q} \exp(\rmi qp) p \nonumber \\
& \times \exp(-\rmi q'p) q' \bra{q'} \frac{\rmd p}{2\pi} \rmd q \rmd q' . 
\end{align}
Next, we employ integration by parts on the integration over $p$ in the second term
\begin{align}
& \int \exp(\rmi qp) p \exp(-\rmi q'p) \frac{\rmd p}{2\pi} \nonumber \\
= & \lim_{t\rightarrow\infty} \left[
\exp(\rmi qp) \frac{p}{-\rmi q'}\exp(-\rmi q'p) \right]_{p=-t}^{p=t} \nonumber \\
& + \int \exp(\rmi qp) \frac{qp}{q'} \exp(-\rmi q'p) \rmd p \nonumber \\
& - \int \exp(\rmi qp) \frac{1}{-\rmi q'} \exp(-\rmi q'p) \rmd p. 
\end{align}
The term with the limit produces a potentially divergent result. Below, we'll consider it separately. Substituting this result back into the full expression, we get
\begin{align}
[\hat{q},\hat{p}] & = \rmi \int \ket{q}\exp(\rmi qp)\exp(-\rmi q'p)\bra{q'}\frac{\rmd p}{2\pi} \rmd q \rmd q' 
- \rmi \hat{\Delta} \nonumber \\
& = \rmi \int \ket{q} \bra{q} \rmd q - \rmi \hat{\Delta}
= \rmi \mathbb{I} - \rmi \hat{\Delta} ,
\end{align}
where
\begin{align}
\hat{\Delta} \eqtri &\ \frac{1}{2\pi} \int \lim_{t\rightarrow\infty} 
[\ket{q} \exp(\rmi qt) t \exp(-\rmi q't) \bra{q'} \nonumber \\
& -\ket{q} \exp(-\rmi qt) (-t) \exp(\rmi q't) \bra{q'}] \rmd q \rmd q'. 
\end{align}
We need to consider the domain of definition of these operators, as determined by the state vectors on which these operators can operate to produce state vectors that are still elements of the same domain. Such state vectors are defined by
\begin{equation}
\ket{\psi} = \int \ket{q} \psi(q) \rmd q ~~~ {\rm and} ~~~
\bra{\psi} = \int \psi^*(q) \bra{q} \rmd q ,
\end{equation}
where $\psi(q)$ and $\psi^*(q)$ are Schwartz functions, as specified for the coefficient space in the definition of the tagged vector space. When we apply $\hat{\Delta}$ on such state vectors, either toward the right $\hat{\Delta}\ket{\psi}$ or toward the left $\bra{\psi}\hat{\Delta}$, we get, for example,
\begin{align}
\hat{\Delta}\ket{\psi} = &\ \frac{1}{2\pi} \int \lim_{t\rightarrow\infty} 
[\ket{q} \exp(\rmi qt) t \exp(-\rmi q't) \braket{q'}{q''} \psi(q'') \nonumber \\
& -\ket{q} \exp(-\rmi qt) (-t) \exp(\rmi q't) \braket{q'}{q''} \psi(q'')] \rmd q \rmd q' \nonumber \\
& = \frac{1}{2\pi} \int \lim_{t\rightarrow\infty} \ket{q} 
\left[\exp(\rmi qt) t \tilde{\psi}(t) \right. \nonumber \\
& \left. + \exp(-\rmi qt) t \tilde{\psi}(-t) \right] \rmd q ,
\end{align}
where
\begin{equation}
\tilde{\psi}(p) = \int \exp(-\rmi q'p) \psi(q') \rmd q'
\end{equation}
is the Fourier transform of $\psi(q')$, and is also a Schwartz function. Therefore,
\begin{equation}
\lim_{t\rightarrow\infty} t \tilde{\psi}(t) = \lim_{t\rightarrow\infty} t \tilde{\psi}(-t) = 0. 
\end{equation}
Likewise for $\bra{\psi}\hat{\Delta}$. Therefore, the terms produced when $\hat{\Delta}$ operates either toward the right or toward the left are zero. We end up with
\begin{equation}
[\hat{q},\hat{p}] = \rmi \mathbb{I} ~~~ {\rm and} ~~~ [\hat{q},\hat{q}] = [\hat{p},\hat{p}] = 0 ,
\label{kommutqp}
\end{equation}
where the commutation of each operator with itself is easily shown to be zero.

Unlike the approach often used in the formulation of quantum theory, these commutation relations are not required to be an assumed postulates. Instead, they follow as a consequence of the postulates for the tagged vector space, particularly from the mutually unbiased postulate. 

\subsection{Ladder operators}

The definition of \emph{ladder operators} (creation and annihilation operators) in terms of tags and extractors are given in terms of the quadrature operators as
\begin{equation}
\hat{a} \eqtri \tfrac{1}{\sqrt{2}}\left(\hat{q}+\rmi\hat{p}\right) ~~~ {\rm and} ~~~
\hat{a}^{\dag} \eqtri \tfrac{1}{\sqrt{2}}\left(\hat{q}-\rmi\hat{p}\right). 
\label{qpnaa}
\end{equation}
Based on (\ref{kommutqp}), the commutation relations of the ladder operators are
\begin{equation}
[\hat{a},\hat{a}^{\dag}] = \mathbb{I} ~~~ {\rm and} ~~~ [\hat{a},\hat{a}] = [\hat{a}^{\dag},\hat{a}^{\dag}] = 0 .
\label{kommutaa0}
\end{equation}

To derive the expressions for ladder operators in terms of tags and extractors, we use identity operators to express both terms with the same types of tags. Considering the annihilation operator, we have
\begin{align}
\hat{a} = &\ \tfrac{1}{\sqrt{2}} \left( \hat{q} + \rmi \hat{p} \right)
= \tfrac{1}{\sqrt{2}} \left( \int \ket{q} q \bra{q}\ \rmd q 
+ \rmi \int \ket{p} p \bra{p}\ \frac{\rmd p}{2\pi} \right) \nonumber \\
= &\ \frac{1}{\sqrt{2}} \int \ket{q_1} \frac{q_1+q_2}{2} \exp(\rmi q_1 p-\rmi q_2 p)
\bra{q_2}\ \frac{\rmd p}{2\pi}\ \rmd q_1\ \rmd q_2 \nonumber \\
& + \rmi\frac{1}{\sqrt{2}} \int \ket{q_1} \exp(\rmi q_1 p) p \exp(-\rmi q_2 p)
\bra{q_2}\ \frac{\rmd p}{2\pi}\ \rmd q_1\ \rmd q_2 \nonumber \\
= & \int \ket{q_1} \frac{1}{\sqrt{2}} \left(\frac{q_1+q_2}{2}+\rmi p\right) \nonumber \\
& \times \exp(\rmi q_1 p-\rmi q_2p) \bra{q_2}\ \frac{\rmd p}{2\pi}\ \rmd q_1\ \rmd q_2 \nonumber \\ 
= & \int \ket{q+\tfrac{1}{2}x} \tfrac{1}{\sqrt{2}} \left(q+\rmi p\right) \nonumber \\
& \times \exp(\rmi x p) \bra{q-\tfrac{1}{2}x}\ \frac{\rmd p}{2\pi}\ \rmd q\ \rmd x .
\label{absweyl}
\end{align}
In the end, we redefined the two $q$-indices as $q_1\rightarrow q+\tfrac{1}{2}x$ and $q_2\rightarrow q-\tfrac{1}{2}x$.  The result comes out to the \emph{Weyl transformation} \cite{weyltra} of a complex quantity
\begin{equation}
\alpha \eqtri \tfrac{1}{\sqrt{2}} \left(q+\rmi p\right) ,
\label{defalph}
\end{equation}
given in terms of the two real unbiased indices serving as the real and imaginary parts, respectively. The inverse transformation
\begin{align}
\alpha & = \tfrac{1}{\sqrt{2}} \left(q+\rmi p\right) \nonumber \\
& = \int \bra{q+\tfrac{1}{2}x}\hat{a}\ket{q-\tfrac{1}{2}x} \exp(-\rmi x p)\ \rmd x ,
\label{wiganni}
\end{align}
identifies the complex quantity as the \emph{Wigner function} \cite{wigner} of the annihilation operator. The creation operator leads to an equivalent representeation, based on the complex conjugation of the complex quantity
\begin{align}
\hat{a}^{\dag} = & \int \ket{q+\tfrac{1}{2}x} \tfrac{1}{\sqrt{2}} \left(q-\rmi p\right) \nonumber \\
& \times \exp(\rmi x p) \bra{q-\tfrac{1}{2}x}\ \frac{\rmd p}{2\pi}\ \rmd q\ \rmd x. 
\label{skepweyl}
\end{align}

\subsection{Phase space}
\label{faseruim}

The expressions in (\ref{absweyl}) and (\ref{wiganni}) deserves careful consideration. The kernels of these operator are combinations of the two unbiased indices \emph{acting as independent variables}. Previously, the unbaised set of tags was defined with the same index space so that all sets of tags are related via unitary transformations. Here, the two unbaised sets of tags are labelled effectively by separate index spaces. If $q\in\mathcal{I}\equiv\mathbb{R}$, then $p\in\tilde{\mathcal{I}}\equiv\mathbb{R}$. The complex quantities $\alpha$ and $\alpha^*$, serving as the kernels of the ladder operaters, are defined on $\mathcal{I}\otimes\tilde{\mathcal{I}}\equiv\mathbb{C}$. Identifying these complex quantities as the Wigner functions of the ladder operator, we find that $\mathcal{I}\otimes\tilde{\mathcal{I}}$ represents a \emph{phase space} \cite{psqm}. It indeed has the properties of a phase space for (one-dimensional) quantum optics in that the commutation relations of the quadrature operators, with indices treated as independent variables, lead to a \emph{symplectic geometry}, which readily follows from (\ref{kommutqp}).

\section{Functional tagged vector spaces}

The development of tagged vector spaces in the context of the Dirac formalism in Section~\ref{appdirac} only provides the formalism for the one-dimensional quantum (particle-number) degree of freedom. By contrast, a formalism that incorporates all the degrees of freedom (spatiotemporal and spin degrees of freedom) is necessarily defined in terms of a 
\emph{function space}. Such a situation is inevitable when the degrees of freedom of the quantum systems become infinite as is often found in physical scenarios of particle physics and quantum optics, among other fields.

Here, we model quantum optics as a tagged vector space in all degrees of freedom. It is accomplished by generalizing the notion of a tagged vector space with a one-dimensional index space to incorporate the spatiotemporal and spin degrees of freedom, using a function space as the index space for the tags and extractions. As a result, the tagged vector space becomes a \emph{functional tagged vector space} in which tags and extractions are functionals.

As suggested by the development in Section~\ref{faseruim}, the elements of the index space can be used to define phase space variables. When the index space is a function space, the phase space variables represent field variables. In the context of quantum optics, they are real \emph{quadrature field variables}, serving as the coordinates of a functional phase space, represented by two \emph{mutually unbiased index spaces}. The functions that are the elements (quadrature field variables) of these index spaces parameterize the spin and spatiotemporal degrees of freedom. These mutually unbiased index function spaces are defined as sets of \emph{real Schwartz functions}, 
so that $\mathcal{I} \eqtri \mathcal{S}(\mathbb{R}^3)$. 

The type of Schwartz functions depends on the nature of the spin degrees of freedom. However, the spin does not introduce complications and can be specified as a generalization of the scalar case. In what follows we'll generally ignore the spin degrees of freedom.

Although one can use Dirac notation to represent the tags and their associated extractors, we retain the original notation for the tags and extractors, representing the tags as $\vlag{q}$ and $\utag{p}$ for the two quadrature field variables, respectively, and their associated extractors are represented by $\ext{q}$ and $\uext{p}$, respectively. The square backets indicate that the tags and extractors are \emph{functionals} of the indices $q$ and $p$, as elements of the function spaces serving as index spaces. The postulates that define the properties of the tags and extractors are then expressed in functional form as
\begin{itemize}
\item \textbf{orthogonality}:
\begin{align}
\begin{split}
\extag{q_1}{q_2} & = \delta[q_1-q_2] ~~~ {\rm and} \\ 
\uexutag{p_1}{p_2} & = (2\pi)^{\Omega}\delta[p_1-p_2] ,
\end{split}
\label{ortax}
\end{align}
where $\delta[\cdot]$ represents a \emph{Dirac delta functional} and $\Omega$ represents the cardinality (the number of degrees of freedom) of the index space,
\item \textbf{completeness}:
\begin{align}
\begin{split}
\int \vlag{q}\ext{q}\ \mathcal{D}[q] & = \mathbb{I} ~~~ {\rm and} \\ 
\int \utag{p}\uext{p}\ \Dcirc[p] & = \mathbb{I} ,
\end{split}
\label{volax}
\end{align}
where $\mathcal{D}[q]$ and $\Dcirc[p]$ are abstract \emph{functional integration measures}, the latter having a factor of $1/(2\pi)$ for each degree of freedom, and $\mathbb{I}$ is the identity operator,
\item \textbf{unbiased}:
\begin{align}
\begin{split}
\exutag{q}{p} & = \exp(\rmi q\diamond p) ~~~ {\rm and} \\ 
\uextag{p}{q} & = \exp(-\rmi q\diamond p) ,
\end{split}
\label{mubax}
\end{align}
where the $\diamond$-contraction is defined as
\begin{equation}
q\diamond p \eqtri \sum_s \int q_s(\mathbf{k})p_s(\mathbf{k})\ \frac{{\rm d}^3 k}{(2\pi)^3} ,
\end{equation}
in terms of the field variables $q_s(\mathbf{k})$ and $p_s(\mathbf{k})$, which are functions of the wave vector $\mathbf{k}$. For the general case, we also show a spin index $s$, although the definition of the Schwartz space that we use below does not incorporate spin.
\end{itemize}
The right-hand sides in (\ref{ortax}) and (\ref{mubax}) are tempered distributions. Expressed in their bare forms, these postulates lead to expressions of linear operators with these distributions acting as Schwartz kernels. The postulates for the $p$-index space incorporate factors of $2\pi$ to anticipate the definition of the Fourier transform.

The coefficient space is
\begin{equation}
\mathcal{C} \eqtri \left\{ F : \mathcal{I} \rightarrow \mathbb{C}, \|F\|=1 \right\} ,
\label{ftagmap}
\end{equation}
a space of normalized \emph{Schwartz functionals}. The norm $\|F\|$ is an Euclidean norm defined as a functional integral.

The functional tagged vector space and the adjoint functional tagged vector space in terms of $q$-tags are
\begin{align}
\begin{split}
\Upsilon & \eqtri \left\{\ket{F} = \int \vlag{q} F[q]\ \mathcal{D}[q] : q\in\mathcal{I}, 
F\in\mathcal{C} \right\} , \\ 
\Upsilon' & \eqtri \left\{\bra{G} = \int G^*[q] \ext{q}\ \mathcal{D}[q] : q\in\mathcal{I} , 
G^*\in\mathcal{C} \right\} .
\end{split}
\end{align}
The equivalent definitions in terms of $p$ produce the same functional tagged vector space and adjoint functional tagged vector space, because the tags that are indexed by $q$ are unitarily related to those indexed by $p$ via the unbiased postulate (\ref{mubax}). This unitary relationship is illustrated by
\begin{align}
\mathbb{I}\ket{F} & = \int \utag{p}\uextag{p}{q} F[q]\ \Dcirc[p]\ \mathcal{D}[q] \nonumber \\
& = \int \utag{p} \exp(-\rmi q\diamond p) F[q]\ \mathcal{D}[q]\ \Dcirc[p] \nonumber \\
& = \int \utag{p} \tilde{F}[p]\ \Dcirc[p] ,
\end{align}
where
\begin{equation}
\tilde{F}[p] = \int \exp(-\rmi q\diamond p) F[q]\ \mathcal{D}[q]
\end{equation}
is the functional Fourier transform. As a result, the elements of the tagged vector space can be represented in terms of tags indexed by either $q$ or $p$ (or any other index unitarily related to $q$) where the coefficient function indexed by $p$ is the Fourier transform of the coefficient function indexed by $q$. Moreover, regardless of the (unitarily related) index used in its definition, the ket $\ket{F}$ always represents the same physical entity.

\subsection{Operators}

In analogy to their definition in Section~\ref{appdirac}, operators on functional tagged vector spaces are generically expressed as
\begin{equation}
\hat{A} \eqtri \int \vlag{\nu} A[\nu,\mu] \ext{\mu}\ \mathcal{D}[\nu,\mu] ,
\label{kwantfop}
\end{equation}
where $\nu$ and $\mu$ represent functions that are unitarily related to functions in the function index space. The kernel $A[\nu,\mu]$ is a general tempered distribution defined on $\mathcal{I}\times\mathcal{I}$. All operators are simultaneously defined on the functional tagged vector space and the adjoint functional tagged vector space, 
as explained in Section~\ref{appdirac}.

The quadrature operators are defined as \emph{first-order moment operators}
\begin{align}
\begin{split}
\hat{q}(\mathbf{k}) & = \int \vlag{q} q(\mathbf{k}) \ext{q}\ \mathcal{D}[q] , \\
\hat{p}(\mathbf{k}) & = \int \utag{p} p(\mathbf{k}) \uext{p}\ \Dcirc[p] .
\end{split}
\label{kwadfdef}
\end{align}
In the functional scenario, they carry wave vector dependencies, inherited from the functions representing the indices. As in Section~\ref{kwadkomm}, one an show that these quadrature operators do not commute, but produce
\begin{equation}
[\hat{q}(\mathbf{k}),\hat{p}(\mathbf{k}')] = \rmi (2\pi)^3 \delta(\mathbf{k}-\mathbf{k}') \mathbb{I} .
\end{equation}

Ladder operators are defined as
\begin{align}
\begin{split}
\hat{a}(\mathbf{k}) & \eqtri \tfrac{1}{\sqrt{2}} \left[\hat{q}(\mathbf{k})+\rmi\hat{p}(\mathbf{k})\right] , \\
\hat{a}^{\dag}(\mathbf{k}) & \eqtri \tfrac{1}{\sqrt{2}} \left[\hat{q}(\mathbf{k})-\rmi\hat{p}(\mathbf{k})\right] ,
\end{split}
\end{align}
leading to commutation relations obtained from those for the quadrature operators
\begin{align}
\begin{split}
[\hat{a}(\mathbf{k}),\hat{a}^{\dag}(\mathbf{k}')] & = (2\pi)^3 \delta(\mathbf{k}-\mathbf{k}') \mathbb{I} , \\
[\hat{a}(\mathbf{k}),\hat{a}(\mathbf{k}')] & = [\hat{a}^{\dag}(\mathbf{k}),\hat{a}^{\dag}(\mathbf{k}')] = 0 .
\end{split}
\end{align}
Using (\ref{kwadfdef}), we obtain expressions for the ladder operators
\begin{align}
\begin{split}
\hat{a} = & \int \vlag{q+\tfrac{1}{2}x} \alpha(\mathbf{k}) \exp(\rmi x\diamond p) \nonumber \\
& \times \ext{q-\tfrac{1}{2}x}\ \mathcal{D}\left[x,q\right]\ \Dcirc[p] , \\
\hat{a}^{\dag} = & \int \vlag{q+\tfrac{1}{2}x} \alpha^*(\mathbf{k}) \exp(\rmi x\diamond p) \nonumber \\
& \times \ext{q-\tfrac{1}{2}x}\ \mathcal{D}\left[x,q\right]\ \Dcirc[p] ,
\end{split}
\end{align}
as \emph{functional Weyl transformations} of complex field variables, given by
\begin{align}
\begin{split}
\alpha(\mathbf{k}) & \eqtri \tfrac{1}{\sqrt{2}} \left[q(\mathbf{k})+\rmi p(\mathbf{k})\right] , \\
\alpha^*(\mathbf{k}) & \eqtri \tfrac{1}{\sqrt{2}} \left[q(\mathbf{k})-\rmi p(\mathbf{k})\right] ,
\end{split}
\end{align}
These complex field variables are expressed in terms of the two mutually unbiased index functions, respectively representing the real and imaginary parts. It leads to a functional phase space representation of operators in terms of \emph{Wigner functionals},
\begin{align}
W_{\hat{A}}[q,p] & = \int \ext{q+\tfrac{1}{2}x}\,\hat{A}\,\vlag{q-\tfrac{1}{2}x} 
\exp(-\rmi x\diamond p)\ \mathcal{D}\left[x\right] \nonumber \\
 & = \int A[q+\tfrac{1}{2}x,q-\tfrac{1}{2}x] \nonumber \\
& \times \exp(-\rmi x\diamond p)\ \mathcal{D}\left[x\right] ,
\end{align}
where the functional kernel $A[q,q']$ defines $\hat{A}$.

\section{Conclusions}

The notion of a tagged vector space provides a structure into which a rigorous mathematical formulation can be embedded in such a way that the resulting formalism conforms to the traditional way in which physical scenarios are modelled. In the context of quantum physics, the tagged vector space formalism allows us to maintain the general usage of Dirac notation and its associated operators, as employed by the physics community. At the same time, it provides a structure in which the rigorous mathematical formulations of quantum theory is incorporated, rendering the tagged vector space formalism mathematically well defined.

While tagged vector spaces are broadly applicable in physics, we focus here on quantum physics as applied in quantum optics to demonstrate its usage. The latter is especially challenging because the combination of the particle-number degree of freedom with spatiotemproal degrees of freedom leads to complex analytical structures seldom found in other fields of physics. Tagged vector spaces are ideally suited to deal with such scenarios.

Historically, the commutation relations emerged from the properties of operators used for observable quantities (position and the derivative with respect to position). The goal in physics is to determine the fundamental mechanisms that would explain phenomena as inevitable consequences. Since operators are mathematical devices used to model processes in quantum physics, their commutation relations are consequences of the properties of these processes and are not regarded as being fundamental. Here, it is demonstrated that they follow as inevitable consequences of the fundamental properties of the physical degrees of freedom. 

On the other hand, canonical commutation relations are often used as a starting point for rigorous mathematical formulations of quantum theory. In view of the presented derivation of these commutation relations, the choice of the set of postulates for an adequate formulation of quantum theory is ambiguous from a purely mathematical point of view. The chosen set of postulates is more a result of the historical development of the current mathematical formulation.

The inner product space of all physical quantum optical states with all degrees of freedom incorporated is represented as a functional tagged vector space with an index space given by a function space (Schwartz space) and a coefficient space represented by all the Schwartz functionals on this function index space. The key ingredient making this formulation possible is the functional integrals in terms of which states and operations are expressed. Such a functional integral can be defined with a probability measure, provided that the integrands are restricted to valid functional probability distributions, or in terms of limits of sequences of such distributions.

The remaining formalism of quantum optics are derived in a straight forward way from what is presented here. Considering the one-dimensional case, one defines a \emph{number operator} in terms of the ladder operators and compute its eigenstates. They are the Fock states and are represented in terms of tags (and extractors) with Hermite-Gauss functions as their coefficient functions. The (right) eigenstates of the annihilation operator (or the left eigenstates of the creation operator) are the coherent states with Schwartz functions as their coefficient functions. Displacement operators and squeezing operators are readily defined as exponentiated operators in terms of ladder or quadrature operators. These developments are performed for an infinite number of degrees of freedom with the aid of the functional approach.

While the expressions in these notes, expressed in terms of tags and extractors, lead to associated expressions for the Fock states, such expressions are obtained from the basic relationships of the ladder operators in standard quantum theory. The same applies for the coherent states. Therefore, the definition of the basic quantum theory in terms of a tagged vector space formalism smoothly matches up with existing quantum optics theory and provides a rigorous mathematical foundation.

We see that the ladder operators are naturally represented in terms of Weyl transforms of complex combinations of the unbiased indices. These complex quantities thus represent the Wigner functions of the ladder operators. The kernels of arbitrary operators, expressed in terms of tags and extractors, are thus related to their Wigner functions. Products of operators leads to Moyal star products of their Wigner functions. The quantum formalism defined in terms of tagged vector spaces naturally leads to a Moyal formalism \cite{groenewold,moyal,psqm}.

\end{document}